\documentclass[conference]{IEEEtran}
\IEEEoverridecommandlockouts
\usepackage{cite}
\usepackage{amsmath,amssymb,amsfonts}
\usepackage{graphicx}
\usepackage{textcomp}
\usepackage{multirow}
\usepackage{xcolor}

\usepackage[ruled,vlined]{algorithm2e}
\usepackage[left=1.62cm,right=1.62cm,top=1.9cm]{geometry}

\usepackage{algpseudocode}
\def\BibTeX{{\rm B\kern-.05em{\sc i\kern-.025em b}\kern-.08em
    T\kern-.1667em\lower.7ex\hbox{E}\kern-.125emX}}
\begin{document}

\title{Improving Multilayer-Perceptron(MLP)-based Network Anomaly Detection with Birch Clustering on CICIDS-2017 Dataset\\

}


\author{
\IEEEauthorblockN{Yuhua Yin\IEEEauthorrefmark{1},
Julian Jang-Jaccard\IEEEauthorrefmark{1},
Fariza Sabrina\IEEEauthorrefmark{2} and
Jin Kwak\IEEEauthorrefmark{3}}
\IEEEauthorblockA{\IEEEauthorrefmark{1}Cybersecurity Lab, Massey University, New Zealand}
\IEEEauthorblockA{\IEEEauthorrefmark{2}School of Engineering and Technology, Central Queensland University, AUSTRALIA}
\IEEEauthorblockA{\IEEEauthorrefmark{3}Department of Cyber Security, Ajou University, REPUBLIC OF KOREA}
yuhua.yin@ieee.org, j.jang-jaccard@massey.ac.nz, f.sabrina@cqu.edu.au, security@ajou.ac.kr }

\maketitle

\begin{abstract}
The network intrusion threats are increasingly severe with the application of computer supported coorperative work. Machine learning algorithms have been widely used in intrusion detection systems, including Multi-layer Perceptron (MLP). In this study, we proposed a two-stage model that combines the Birch clustering algorithm and MLP classifier to improve the performance of network anomaly multi-classification. In our proposed method, we first apply Birch or Kmeans as an unsupervised clustering algorithm to the CICIDS-2017 dataset to pre-group the data. The generated pseudo-label is then added as an additional feature to the training of the MLP-based classifier. The experimental results show that using Birch and K-Means clustering for data pre-grouping can improve intrusion detection system performance. Our method can achieve 99.73\% accuracy in multi-classification using Birch clustering, which is better than similar researches using a stand-alone MLP model.  \\

\end{abstract}


\section{Introduction}
Organizations and individuals face increasing threats of cyber-attacks while enjoying the convenience of computer supported coorperative work. According to a report by Check Point Research, cyber-attacks in 2021 have increased by 50\%, and each organization would experience approximately 900 attacks per week\cite{b5}. In addition, the types of network intrusions have become more diverse, and more attacks come from zero-day attacks\cite{b6}. Network intrusion detection systems are designed to identify network attacks by analyzing normal and malicious behavior of network traffic\cite{b7}. Traditional signature-based intrusion detection systems have been unable to deal with complex and diverse modern network attacks, so increasing research has focused on anomaly detection-based methods. Machine learning models have been widely used in anomaly-based intrusion detection systems and have good detection performance\cite{b7}. Supervised learning and unsupervised learning are two common machine learning methodologies. Models based on supervised learning, such as MLP, decision tree, random forest, SVM, etc., can learn and classify labeled data\cite{b8}. Models based on unsupervised learning, including Kmeans, Birch, hidden Markov models, etc., can cluster similar unlabeled data to find anomalies\cite{b9}. Furthermore, datasets used for intrusion detection are essential and can be divided into packet-based and flow-based data\cite{b10}. The packet-based dataset mainly collects some meta-information such as protocol and size in the packet. In contrast, the flow-based dataset treats the packets that share attributes within a specific time window as a flow. Many modern intrusion detection datasets are flow-based, such as CICIDS-2017, CICIDS-2018, etc.\cite{b10}. The timeliness of intrusion detection datasets also affects their effectiveness. Early datasets such as NSL-KDD only contain some simple network attack classes\cite{b11}. With the continuous emergence of network attack types and variants, newer datasets such as CICIDS-2017 can better reflect modern and complex network anomalies\cite{b12}.

This paper proposed a network anomaly detection method that combines both supervised and unsupervised learning models. In our method, the IDS data is first pre-labelled using the Birch clustering algorithm, and then the generated pseudo-label is used as an extra feature in the training of the MLP classifier. The CICIDS-2017 as a modern IDS dataset was used to validate our method.

The contributions of our research are as follows:
\begin{itemize}
\item 
We proposed an approach that can improve MLP-based intrusion detection systems by using an unsupervised clustering algorithm for pre-labeling, which is believed to help the classifier distinguish between similar samples.
 
\item 
During data pre-processing of CICIDS-2017, we removed duplicate data, regrouped the labels based on recent research, and resampled benign traffic to make benign samples and attack samples equal. Also, we applied information gain as a feature selection method to remove 18 unimportant features.
 
\item 
Our experimental results demonstrate that our model can achieve a multi-class accuracy of 99.73\%, which is better than using MLP only. After comparison, our method outperformed similar research.

\end{itemize}

We organized the rest of the paper as follows. In Section~\ref{sec:rw}, we reviewed related works on machine learning-based intrusion detection systems.   In Section~\ref{sec:pr} we described the background of our proposed method. In Section~\ref{sec:pm}, we introduced our proposed method. In Section~\ref{sec:re}, we described our experiment process and demonstrated the results. Finally, in Section~\ref{sec:con}, we concluded our work and discussed our future works.

\section{Related Work}
\label{sec:rw}
There have been many studies using different machine learning methods and IDS datasets for network anomaly detection. In this section, we reviewed some related work.

Wen et al. proposed an autoencoder-based network anomaly detection method in the NSL-KDD dataset\cite{b11}. The researchers used the percentile method to remove outliers before applying the autoencoder, and then implemented a five-layer autoencoder model for classification. The proposed autoencoder model can achieve a state-of-art accuracy of 90.61\% on the multi-classification task for NSL-KDD.

Kasongo and Sun proposed a feature selection method using XGBoost as an intrusion detection dataset and applied it to the UNSW-NB15 dataset\cite{b13}. In the experiments, multiple machine learning methods including SVM, KNN, LR, ANN and DT are used to verify the performance of the intrusion detection system. The results show that the feature selection method of XGBoost enables the decision tree to obtain the highest binary classification performance up to 90\% in UNSW-NB15.

Jing and Chen proposed a network intrusion detection method based on SVM\cite{b14}. A new data scaling method based on the log function is used, which is different from the traditional min-max scaling. Experimental results show that the proposed method can achieve 85.99\% binary classification accuracy on UNSW-NB15.

Khan et al. proposed a novel two-stage network anomaly detection method\cite{b15}. Their method is divided into two steps, first using an autoencoder to perform binary classification on the data and then adding the obtained binary classification result as an extra feature to the multi-classification in the second step. The method achieves 99.99\% and 89.13\% multi-class accuracy on KDD99 and UNSW-NB15, respectively.

Rosay et al's study used MLP as a classifier to validate their intrusion detection method on CICIDS-2017 and CICIDS-2018\cite{b16}. The proposed MLP model consists of two hidden layers, each containing 256 neurons. The authors removed constant-valued features from the datasets. Compared with the studies that also used MLP as a classifier, the experiments achieved better multi-classification accuracy of 99.46\% and 95.47\% on CICIDS-2017 and CICIDS-2018.

Ho et al. explored resampling methods for intrusion detection systems using CICIDS-2017\cite{b17}. Random undersampling, SMOTE, and their combination were used with the C4.5 classifier separately in the experiment. The results show that the best performance is obtained using random undersampling.

In Ustebay et al.'s study, recursive feature elimination(RFE) was used as a feature selection method for intrusion detection\cite{b18}. On the CICIDS-2017 dataset, the researchers used RFE to obtain the ten most important features, which were then used in the Deep Multilayer Perceptron(DMLP) classifier. The experiment achieved 91\% accuracy.

Kong et al. proposed an intrusion detection system using a hybrid deep learning approach combining 1D CNN and LSTM\cite{b19}. In the proposed method, 1D CNN is first used to extract spatial features.Then the connected LSTM can extract temporal features, so that the model can extract both temporal and spatial features simultaneously. The results of the experiments can achieve higher than 97\% accuracy on CICIDS-2017.

\section{Preliminaries}
\label{sec:pr}

\subsection{CICIDS2017 Dataset}

CICIDS-2017 is a flow-based intrusion detection dataset created by the Canadian Institute for Cyber Security (CIC) to solve the problems of previous datasets\cite{b1}. The dataset recorded network traffic for five days and contained 2,830,743 samples consisting of 15 classes and 78 features. Among 15 classes, there were benign instances and 14 types of attack instances. The number of each class is shown in Table \ref{tab:cicnum}. The 78 features are all integer or float numeric features.

\begin{table}[]
\centering
\caption{Records of Original CICIDS-2017 Dataset}
\label{tab:cicnum}
\begin{tabular}{ll}
\hline
\textbf{Label}             & \textbf{Instances} \\
\hline
BENIGN                     & 2273097         \\
DoS Hulk                   & 231073          \\
PortScan                   & 158930          \\
DDoS                       & 128027          \\
DoS GoldenEye              & 10293           \\
FTP-Patator                & 7938            \\
SSH-Patator                & 5897            \\
DoS slowloris              & 5796            \\
DoS Slowhttptest           & 5499            \\
Bot                        & 1966            \\
Web Attack - Brute Force   & 1507            \\
Web Attack - XSS           & 652             \\
Infiltration               & 36              \\
Web Attack - Sql Injection & 21              \\
Heartbleed                 & 11          \\
\textbf{Total} & 2830743\\
\hline
\end{tabular}
\end{table}

\subsection{Birch Clustering Algorithm}
Birch (balanced iterative reducing and clustering using hierarchies) is an efficient unsupervised hierarchical clustering algorithm proposed by Zhang et al.\cite{b20}. Birch is suitable for large datasets and only needs to read the data once to cluster the data. Birch constructs a CF tree data structure and uses the Clustering Feature at each leaf node to represent a sub-cluster by compressing instances and features. Each Clustering Feature consists of a vector with three elements, as shown in equation \ref{e:birch}. Birch clustering algorithm usually has three important parameters: $threshold$, $branching\_factor$, and $n\_clusters$. $Threshold$ determines the radius threshold of the largest sample in a single leaf node. $branching\_factor$ determines the largest number of samples in a node, and $n\_clusters$ represents the final number of clusters to decide the aggregation of nodes.

\begin{equation} \label{e:birch}
   CF=(N,LS,SS)
\end{equation}
where N represents the number of data points, LS represents the linear sum of N data points, and SS represents the square sum of N data points.

\subsection{K-Means Clustering Algorithm}

K-means is a centroid-based unsupervised clustering algorithm that is efficient and easy to implement. In the K-means algorithm, k centroids are randomly initialized, and then the samples are grouped according to the minimum euclidean distance between the data points and the centroids. After that, new centroids are calculated based on the mean of the data points in each cluster. This process runs iteratively until each cluster is converged and no longer changes.

\begin{algorithm}
		\SetAlgoLined
		\DontPrintSemicolon
		\label{RFE}
		
		\KwIn{
		    
		    D: data points $\{d_1,d_2,\ldots,d_m\}$\\
		    K: centroids $\{k_1,k_2,\ldots,k_n\}$
		}
		\KwOut{
		    
			C: clustering result $\{c_1,c_2,\ldots,c_m\}$
		}
		
        \Begin
		{
		    initialize centroids for $\{k_1,k_2,\ldots,k_n\}$\\
		    \Repeat{centroids do not change}{
		        assign clusters for $\{c_1,c_2,\ldots,c_n\}$ by calculating the minimum euclidean distance between data points and each centroid.\\ 
		        calculate new centroids by cluster.\\
		    }
			
			}
		\caption{K-means algorithm} 
\end{algorithm}

\subsection{Multi-layer Perceptron Classifier}
\begin{figure}
\centering
\includegraphics[width=0.48\textwidth]{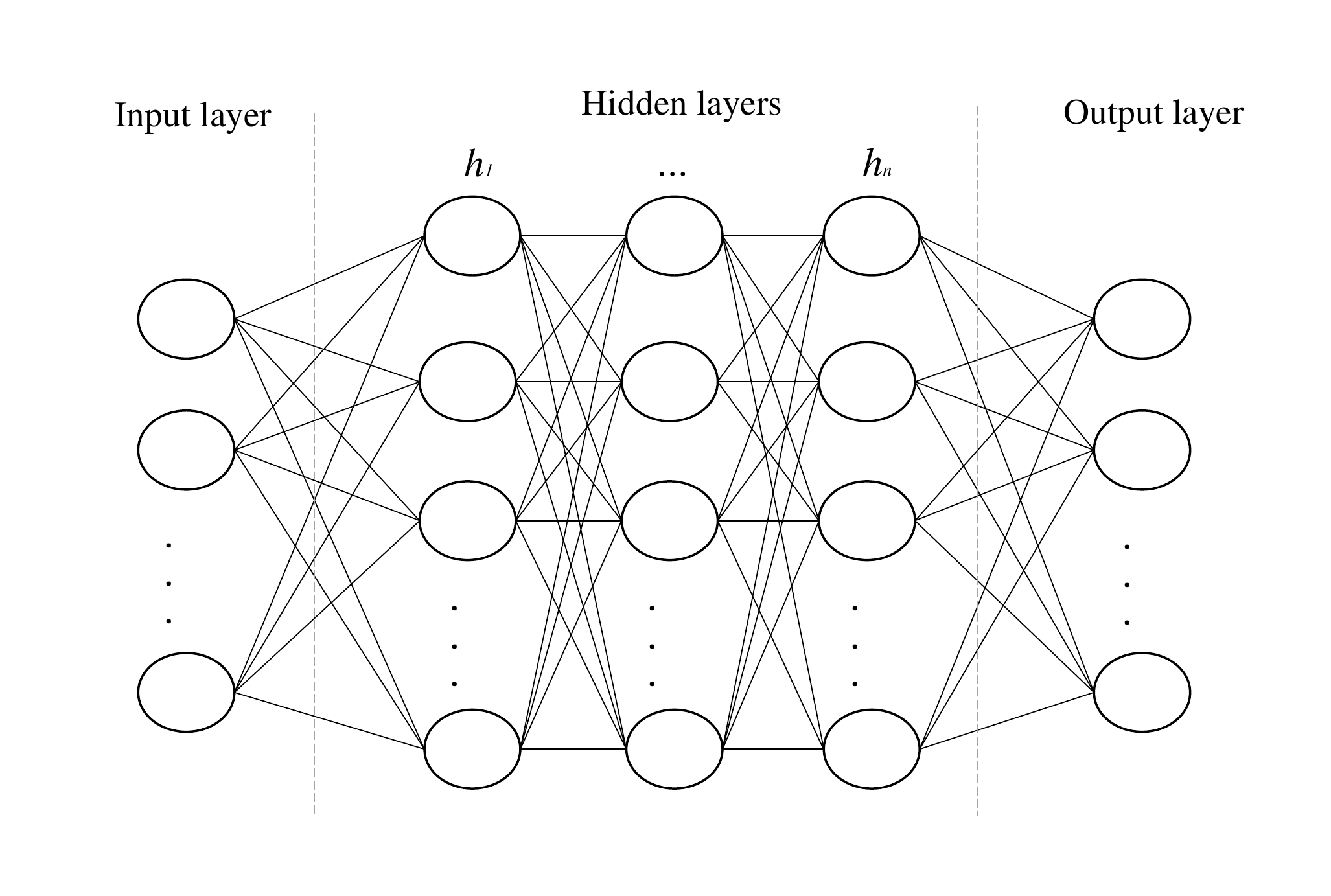}
\caption{Basic MLP Model}
\label{fig:figure1} 
\end{figure}

As seen in Figure \ref{fig:figure1}, Multi-layer Perceptron (MLP) is a feed-forward artificial neural networks composed of multiple layers with activation functions. An MLP typically consists of at least three layers including an input layer, a hidden layer, and an output layer. All layers are fully connected and use a supervised backpropagation for training. When an MLP is used for a classification task, it creates the same number of neurons in the input layer according to the number of features while the number of neurons in the last output layer is decided by the number of classes to be classified. The output is calculated by computing weighted inputs and a bias associated with the layer (shown in \ref{e:e1}). 

\begin{equation} \label{e:e1}
   Z^{[l]}=W^{[l]}A^{[l-1]}+b^{[l]}
\end{equation}
where W\textsuperscript{[l]} indicates the weight inputs, b\textsuperscript{[l]} depicts a bias, and Z\textsuperscript{[l]} is the output.

An activation function is used at each layer to normalize the output in a certain range (i.e., typically between -1 and 1 and its variations) to improve computation efficiency (shown in equation \ref{e:e2}).
\begin{equation} \label{e:e2}
   A^{[l]}=g\left ( Z^{[l]} \right )
\end{equation}
where Z\textsuperscript{[l]} is the output, $g$ represents an activation function, and A\textsuperscript{[l]} is the activated result.

The error between the value predicted by the model and the original value is calculated by a loss function as shown in equation \ref{e:e3}. The result of the loss function at each interaction is used by the supervised backpropagation mechanism to update the weights associated with each input in the layer and a bias.

\begin{equation}\label{e:e3}
	L(y,\hat{y}) = \frac{1}{m}\sum_{i=1}^{m}(y{_i}-\hat{y{_i}})^2
\end{equation}
where m indicates the total number of input samples,$\hat{y}$ represents the predicted value by the model, and y is the original value.

\subsection{Information Gain}
We used Information Gain(IG) as a filter-based feature selection technique to select the most relevant subset of available features in a dataset to reduce the noise from the data and improve the classification accuracy.

The underlying IG utilizes information entropy which measures the disorder of the system. A system with high entropy is considered unpredictable and more disordered while low entropy is associated with highly predictable and less disorderly. In machine learning training, information entropy is used to measure the level of predictability of data distribution. For example, low entropy means many similar values in data distribution while high entropy indicates a lot of different values.

Mathematically, entropy is defined as follows:
\begin{equation} \label{e:e5}
    H\left( Y \right) = - \sum\nolimits_{i = 1}^n {p\left( {{y_i}} \right)lo{g_2}p\left( {{y_i}} \right)}
\end{equation}
where, $n$ is the number of classes in the dataset $Y$ and $p(y\textsubscript{i})$ indicates the probability of picking an element $y\textsubscript{i}$ in each class of the dataset $Y$.

The average specific conditional entropy (i.e., another entropy value given that we already know the entropy of $X$) can be defined as follows:
\begin{equation} \label{e:e6}
    H\left( {Y|X} \right) = - \sum\nolimits_{i = 1}^m {p\left( {{x_i}} \right)H\left( {Y|X = {x_i}} \right)}
\end{equation}
where, $m$ is the number of classes in the dataset $X$ and $p(x\textsubscript{i})$ indcates the probability of picking an element $x\textsubscript{i}$ in each class of the dataset $X$.

In a simple term, IG can be seen as the amount of entropy removed. Mathematically, this can be defined as follows:
\begin{equation} \label{e:e7}
    IG\left( {Y,X} \right) = H\left( Y \right) - H\left( {Y|X} \right)
\end{equation}

In summary, the higher the IG, the more entropy is removed, and the more information the dataset $Y$ carries about $X$. After calculating the IGs for all different features in the input, we rank them and choose the top $n$ features to feed to the MLP model.


\section{Proposed Method}
\label{sec:pm}

\subsection{Overview}
\begin{figure*}
\centering
\includegraphics[width=0.98\textwidth]{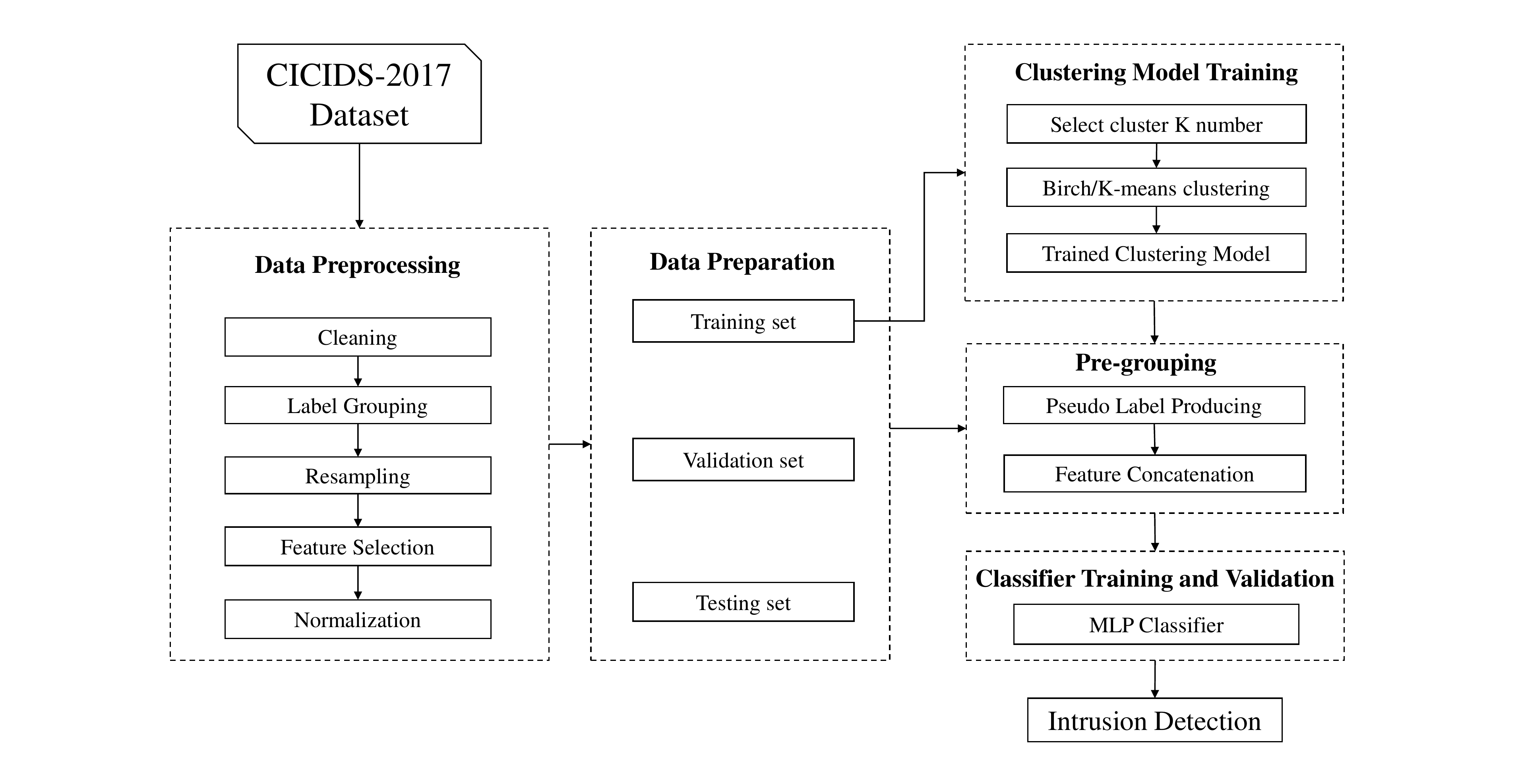}
\caption{Our Proposed Method}
\label{fig:method} 
\end{figure*}

This section described the workflow of our proposed two-stage intrusion detection method. The CICIDS-2017 dataset provides 2.8 million instances of data with 78 numerical features, and they need to be pre-processed at first. In the data pre-processing step, we performed cleaning, label grouping, resampling, information gain feature selection, and normalization on the data (see Figure \ref{fig:method}). We divided the pre-processed data into a training set, a validation set, and a test set according to the 80:10:10 proportion, which is unseen to each other. Our method mainly involved two machine learning modules, an unsupervised clustering model and an MLP deep learning classifier. Only training data is used to train the Birch or K-Means unsupervised clustering algorithm because the training data is assumed to be known. We used the elbow method when choosing the cluster number k, which was described in detail in the following subsection. After obtaining a clustering model, we generated cluster labels for the training, validation ,and test set. This pseudo-label is added as an additional feature to the MLP model along with the original features. Finally, we will use the test set and the trained MLP model to verify the effectiveness of our method.

\subsection{Elbow Method to Select Cluster Number K}
The elbow method is a heuristic cluster number k selection method applied to unsupervised clustering algorithms with a cluster number parameter\cite{b25}. As the number of clusters increases, the mutual information within each cluster will be higher, but too many clusters can cause overfitting problems. The Elbow method selects the inflection points from the cluster purity line graph, which can help us choose a potentially suitable cluster number before overfitting. In our method, we used information gain to measure the clustering performance by calculating the entropy between the generated cluster label and the actual class label of the training set. After plotting the information gain metric under different cluster numbers, we can decide potential cluster numbers at the elbow points to test our model.

\subsection{Specified MLP Classifier}

\begin{figure}
\centering
\includegraphics[width=0.48\textwidth]{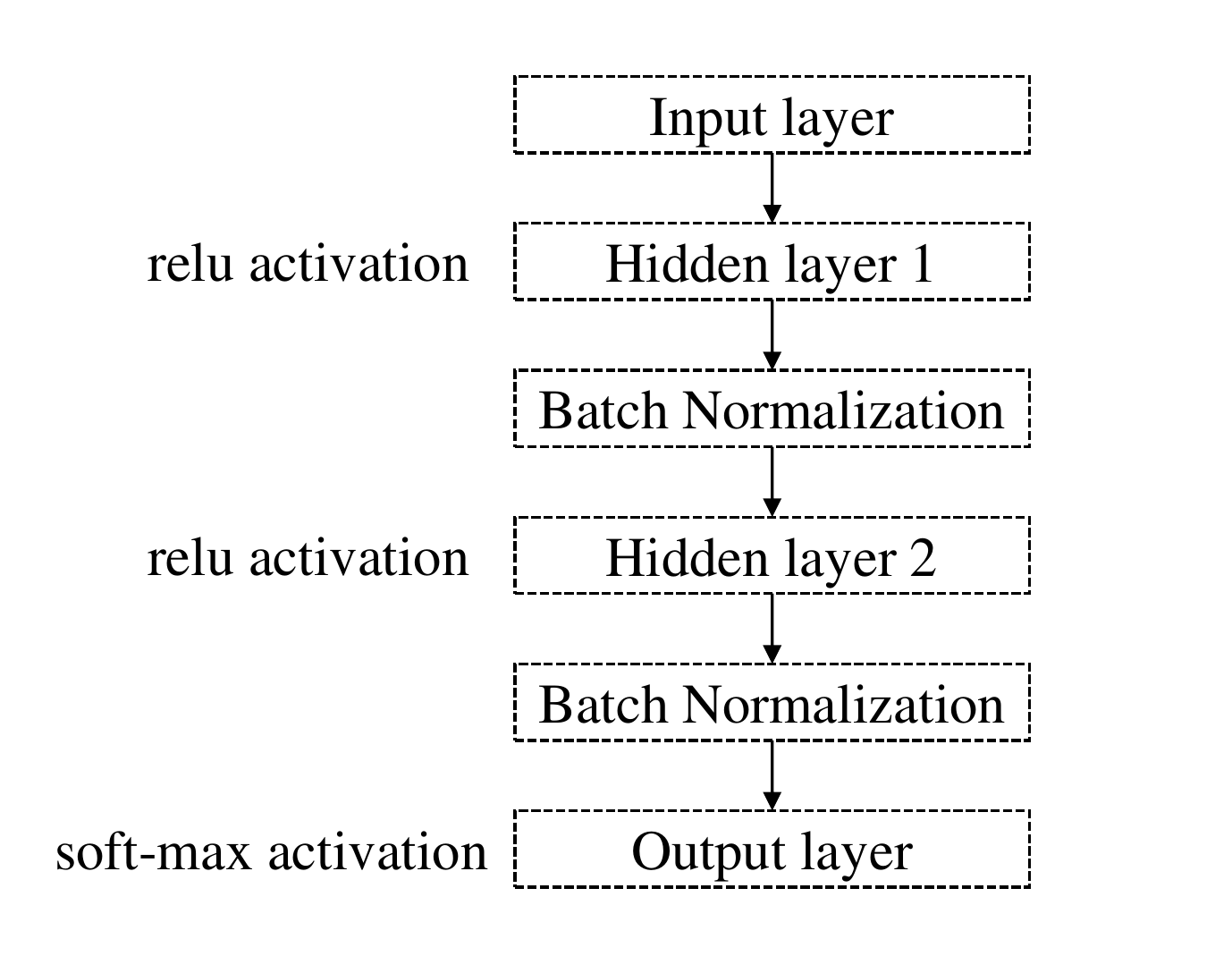}
\caption{Our Specified MLP Classifier}
\label{fig:ourmlp} 
\end{figure}


\begin{table}[]
\centering
\caption{Hyper-Parameter setting for our MLP model}
\label{tab:hyper2}
\begin{tabular}{ll}
\hline
\textbf{Parameter}         & \textbf{Setting} \\
\hline
neurons for hidden layer 1 & 256              \\
neurons for hidden layer 2 & 256              \\
output activation          & softmax          \\
optimizer                  & adam             \\
learning rate              & 0.0001           \\
batch size                 & 64               \\
epochs                     & 200              \\
early stopping patient     & 20       \\
\hline
\end{tabular}
\end{table}

We implemented an MLP model with two hidden layers as the classifier in our proposed method. As shown in Figure \ref{fig:ourmlp}, each hidden layer contains 256 neurons and uses relu activation. Batch Normalization is added after each hidden layer for regularization, and it helps the model to avoid overfitting caused by gradient vanishing and gradient exploding. In the output layer, softmax activation is used to output a normalized probability vector for each class. The final classification result uses the argmax function to obtain the class with the highest probability. In addition, the Adam optimization algorithm is used in our MLP model to adjust the learning rate automatically. The early-stopping technique is used to prevent overfitting. If the validation loss do not continue to decrease in the specified patient's epochs, the training can be terminated early. The relevant hyperparameter settings were listed in Table \ref{tab:hyper2}.
\section{Experiments and Results}
\label{sec:re}

\subsection{Hardware and Environment Setting}

Our research was conducted on a desktop computer running Ubuntu 20.04.4 LTS. The Desktop computer is equipped with 16GB of RAM, a Ryzen 2700 processor, as well as an RX580 graphics card. TensorFlow 2.4.1 was utilized to develop the MLP model in our Python 3.8-based experimental environment. For our work, Scikit-Learn, Numpy, pandas, and matplotlib supplied pre-processing, feature selection, and visualization functions. Table \ref{table:hard} details hardware and environmental parameters.


\begin{table}[]
\centering
\caption{Hardware and Environment Specification}
\label{table:hard}
\begin{tabular}{ll}
\hline
\textbf{Unit}    & \textbf{Description}       \\
\hline
Processor        & AMD Ryzen 7 2700                                                                                         \\
RAM              & 16 GB                                                                                                    \\
GPU              & AMD RX580                                                                                                \\
Operating System & Ubuntu 20.04.4 LTS                                                                                       \\
Packages         & \begin{tabular}[c]{@{}l@{}}Tensorflow 2.4.1, Sklearn 1.0.2, \\ Numpy, Pandas and Matplotlib\end{tabular} \\
\hline
\end{tabular}
\end{table}

\subsection{Data Pre-processing}
The procedure and methods we applied for data pre-processing were described in this subsection.
\subsubsection{Cleaning}
We eliminated 2867 rows with null values from the CICIDS-2017 dataset, as well as the Label column, which specified the class type. 

\subsubsection{Label Grouping}

There are 15 labels for instances in the original CICIDS-2017 dataset, some of which can be grouped. To help reduce the output vector of classification problems, Panigrahi et al \cite{b2} and Kurniabudi et al \cite{b3} proposed a grouping method that can group 15 labels into 7 classes. We described the labels before and after grouping in Table \ref{tab:my-table1}.

\begin{table}[]
\centering
\caption{Grouping labels for CICIDS-2017}
\label{tab:my-table1}
\begin{tabular}{lll}
\hline
\textbf{New Labels} & \textbf{Original Labels}                                                                                                  & \textbf{Number}  \\
\hline
Benign              & Bengin                                                                                                                    & 2271320          \\
Bot                 & Bot                                                                                                                       & 1956           \\
Brute Force         & \begin{tabular}[c]{@{}l@{}}FTP-Patator,\\ SSH-Patator\end{tabular}                                                        & 13832           \\
DoS/DDoS            & \begin{tabular}[c]{@{}l@{}}DDoS, DoS, GoldenEye, DoS Hulk,\\ DoS Slow, httptest, DoS slowloris,\\ Heartbleed\end{tabular} & 379748            \\
Infiltration        & Infiltration                                                                                                              & 36             \\
PortScan            & PortScan                                                                                                                  & 158804             \\
Web Attack          & \begin{tabular}[c]{@{}l@{}}Web Attack-Brute Force,\\ Web Attack-Sql Injection, Web Attack-XSS\end{tabular}                & 2180               \\
\textbf{Total}      &                                                                                                                           & \textbf{2827876}\\
\hline
\end{tabular}
\end{table}

\subsubsection{Resampling}

\begin{figure}
    \centering
    \includegraphics[width=0.24\textwidth]{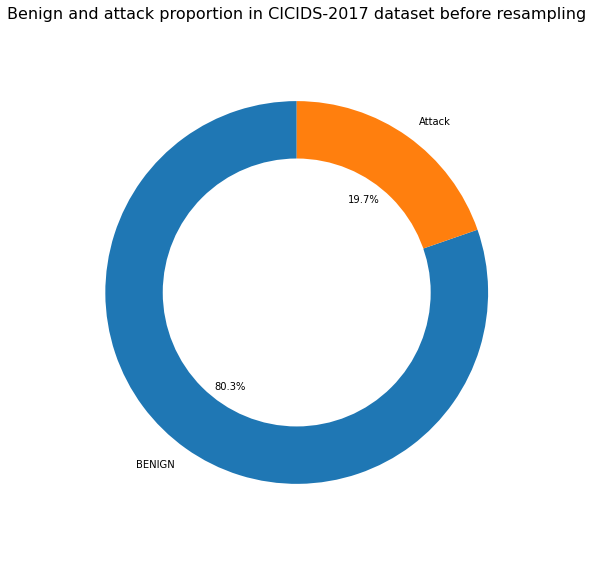} 
    \includegraphics[width=0.235\textwidth]{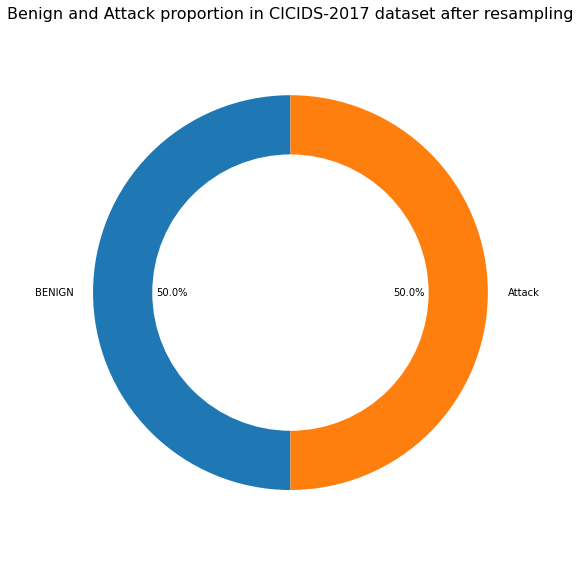}

    \caption{Resampling benign samples in CICIDS-2017}
    \label{Over2}
\end{figure}

In the original CICIDS-2017 dataset, benign samples account for almost 80\% of the total 2.8 million samples, which is imbalanced. We used the random undersampling method which randomly remove majority instances on benign instances and made the benign and attack samples equal (See Figure \ref{Over2}).

\subsubsection{Feature Selection}

\begin{figure}
\centering
\includegraphics[width=0.48\textwidth]{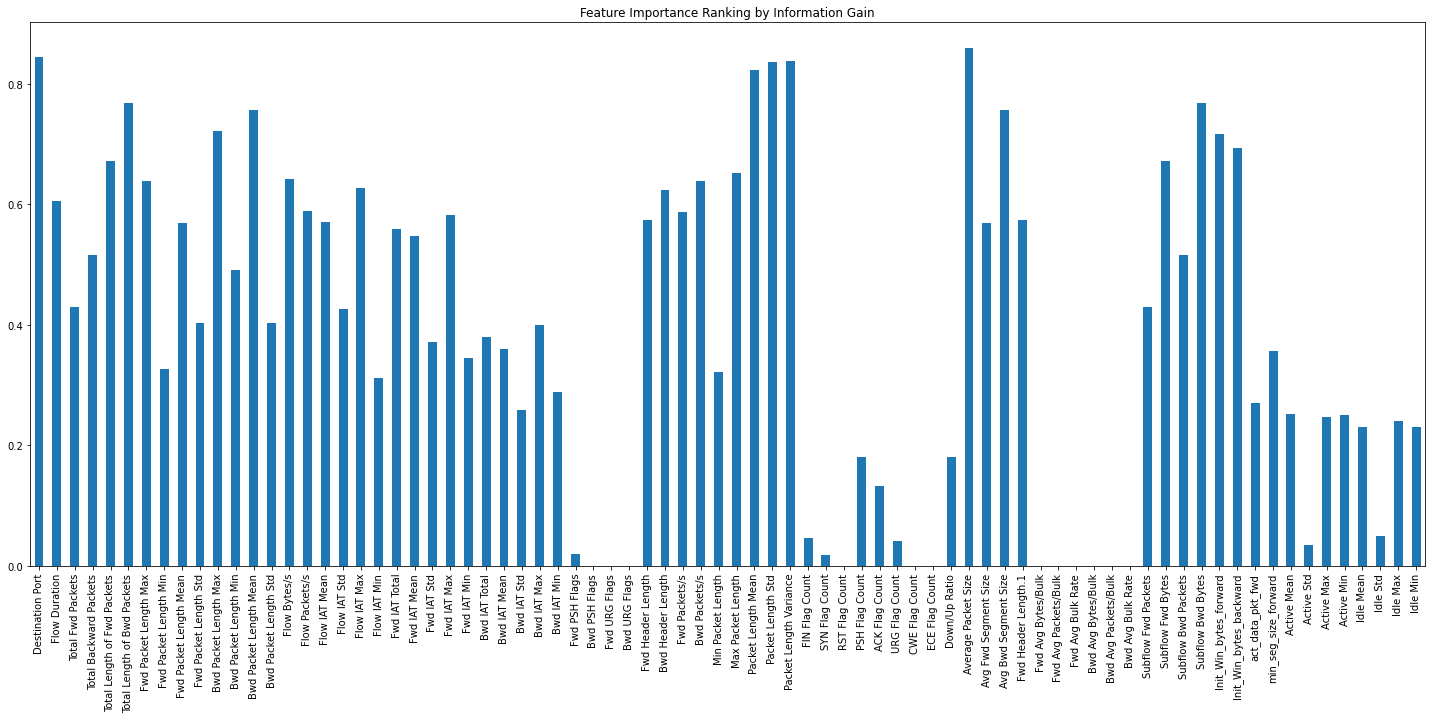}
\caption{Information Gain Feature Importance Ranking}
\label{fig:feature} 
\end{figure}

Figure \ref{fig:feature} depicted the ranking results after we calculated the importance of each feature using Information Gain. The figure shows that the importance scores of some features are very low, implying that these features are either irrelevant or redundant. We removed 18 features with information gain importance less than 0.1, leaving 60 features for our model.

\subsubsection{Normalization}

Normalization can unify the value range of each feature and remove bias produced by disparate value scales during MLP model training. To transform the range of feature values between 0 and 1, we utilized MinMax Normalization\cite{b4}. The new value is computed by dividing the difference between the min and max values by the scale size, as depicted in equation \ref{minmax}.

\begin{equation}
x\textsubscript{$i$}'=\frac {x\textsubscript{$i$}-\min(x\textsubscript{$i$}) }{\max(x\textsubscript{$i$}) -\min(x\textsubscript{$i$}) }
\label{minmax}
\end{equation}
where $x\textsubscript{i}$ depicts all features, min($x\textsubscript{i}$) is the minimum value among all the features, and max($x\textsubscript{i}$) is the the maximum value among all the features.

\subsubsection{Training, validation and test data preparation}

\begin{figure}
    \centering
    \includegraphics[width=0.25\textwidth]{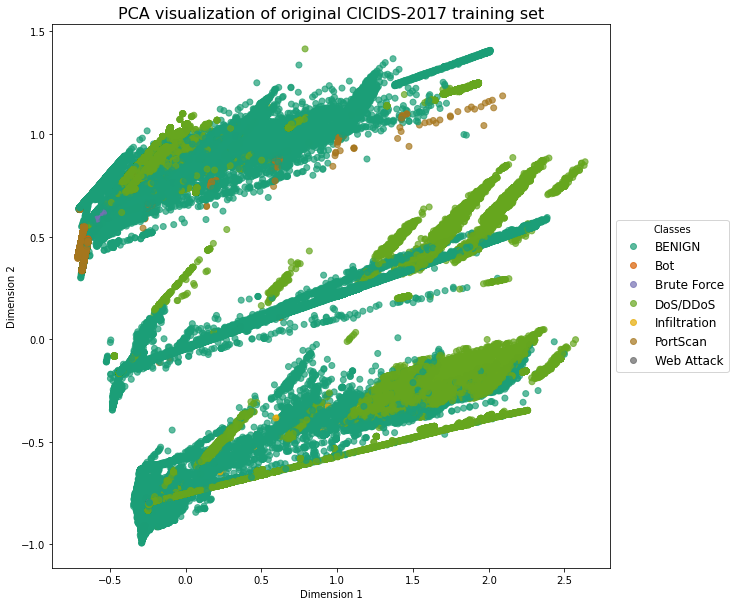} 
    \includegraphics[width=0.21\textwidth]{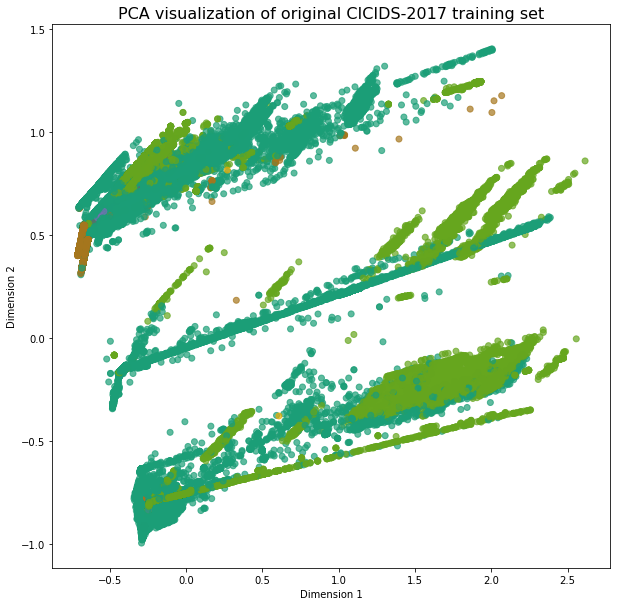}
    \caption{The PCA visualization of training and test set for CICIDS-2017}
    \label{PCA2}
\end{figure}

In the ratio of 80:10:10, we set aside a training set, validation set, and test set from the resampled dataset for the CICIDS-2017 dataset. The number of occurrences is shown in Table \ref{tab:records}. Figure \ref{PCA2} also displays the hold-out training and test set shown using PCA.

\begin{table}[]
\centering
\caption{Prepared training, validation and test dataset}

\label{tab:records}
\begin{tabular}{llll}
\hline
\textbf{Label} & \textbf{Training set} & \textbf{Validation set} & \textbf{Test set} \\
\hline
Benign         & 445244                & 55656                   & 55656             \\
DoS/DDoS       & 303798                & 37975                   & 37975             \\
PortScan       & 127043                & 15881                   & 15880             \\
Brute Force    & 303798                & 1383                    & 1383              \\
Web Attack     & 1744                  & 218                     & 218               \\
Bot            & 1565                  & 196                     & 195               \\
Infiltration   & 29                    & 3                       & 4                 \\
\textbf{Total} & 890489                & 111312                  & 111311           \\
\hline
\end{tabular}
\end{table}

\begin{table*}[]
\centering
\caption{Results of Different Cluster Numbers under Clustering Algorithms}
\label{tab:k_result}
\begin{tabular}{llllll}
\hline
\textbf{Method}    & \textbf{K(Cluster) number} & \textbf{Precision(\%)} & \textbf{Recall(\%)} & \textbf{F1 score(\%)} & \textbf{Accuracy(\%)} \\
\hline
MLP(78 features)   & -                          & 99.01                  & 99.01               & 98.94                 & 99.01                 \\
MLP(60 features)   & -                          & 99.45                  & 99.45               & 99.44                 & 99.45                 \\
K-means+MLP        & 3                          & 99.68                  & 99.69               & 99.68                 & 99.69                 \\
K-means+MLP        & 4                          & 99.70                  & 99.70               & 99.69                 & 99.70                 \\
K-means+MLP        & 6                          & 99.41                  & 99.41               & 99.41                 & 99.41                 \\
K-means+MLP        & 8                          & 99.61                  & 99.60               & 99.60                 & 99.60                 \\
Birch+MLP          & 3                          & 99.59                  & 99.59               & 99.57                 & 99.59                 \\
Birch+MLP          & 9                          & 99.69                  & 99.69               & 99.69                 & 99.69                 \\
\textbf{Birch+MLP} & \textbf{12}                & \textbf{99.73}         & \textbf{99.73}      & \textbf{99.73}        & \textbf{99.73}     \\
\hline
\end{tabular}
\end{table*}

\begin{table}[]
\centering
\caption{Evaluation Metrics of Multi-classification for CICIDS-2017}
\label{tab:result}
\begin{tabular}{lllllc}
\hline
             & \textbf{Precision} & \textbf{Recall} & \textbf{F1 score} & \textbf{FPR} & \multicolumn{1}{l}{\textbf{Accuracy}} \\
\hline
Benign       & 0.9985             & 0.9965          & 0.9975            & 0.0015       & \multirow{8}{*}{\textbf{99.73\%}}     \\
Bot          & 0.9034             & 0.6718          & 0.7706            & 0.0001       &                                       \\
Brute Force  & 0.9788             & 1.0000          & 0.9893            & 0.0003       &                                       \\
DoS/DDoS     & 0.9962             & 0.9989          & 0.9980            & 0.0020       &                                       \\
Infiltration & 1.0000             & 0.7500          & 0.8571            & 0.0000       &                                       \\
PortScan     & 0.9986             & 0.9994          & 0.9990            & 0.0002       &                                       \\
Web Attack   & 0.9904             & 0.9450          & 0.9671            & 0.0000       &                                       \\
Avg.         & 0.9973             & 0.9973          & 0.9973            & 0.0015       &         \\
\hline
\end{tabular}
\end{table}

\begin{figure}
    \centering
    \includegraphics[width=0.235\textwidth]{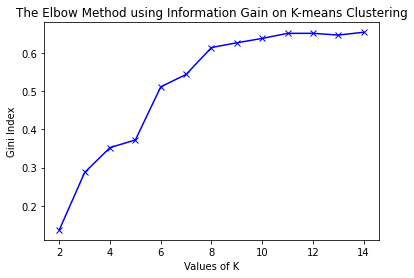} 
    \includegraphics[width=0.235\textwidth]{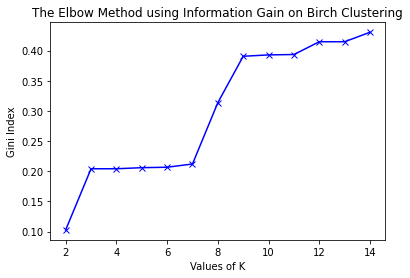}
    \caption{The Elbow Method using Birch and K-means}
    \label{elbow}
\end{figure}

\begin{table*}[]
\centering
\caption{Comparison of Similar Work}
\label{tab:compare}
\begin{tabular}{lllllll}
\hline
\textbf{Work}                 & \textbf{Model}     & \textbf{Precision(\%)} & \textbf{Recall(\%)} & \textbf{FPR(\%)}  & \textbf{F1 score(\%)} & \textbf{Accuracy(\%)} \\
\hline
Rosay et al.\cite{b16}         & MLP                & 99.51              & 99.41           & 0.49          & 99.46             & 99.46             \\
Ustebay et al. \cite{b18}      & DMLP               & -                  & -               & -             & -                 & 91                \\
Jiang et al. \cite{b26}        & MLP                & 99.87              & 99.60           & -             & 99.41             & 99.23             \\
Jabbar and Mohammed \cite{b27} & MLP                & -                  & -               & -             & -                 & 98.98             \\
Azzaoui et al. \cite{b28}      & DNN                & 80.33              & -               & 0.07          & -                 & 99.43             \\
Alrowaily et al. \cite{b29}    & MLP                & 95.36              & 96.25           & -             & 95.57             & 96.26             \\
\textbf{Proposed Method}      & \textbf{Birch+MLP} & \textbf{99.73}     & \textbf{99.73}  & \textbf{0.15} & \textbf{99.73}    & \textbf{99.73}   \\
\hline
\end{tabular}
\end{table*}
In our experiments, the elbow method was used to select the cluster number k of the Birch and K-means clustering algorithm, and information gain was used as the metric of the cluster purity. A higher information gain metric means a better cluster purity. Default hyper-parameter $Threshold$=0.5 and $branching\_factor$=50 in sklearn were used for Birch algorithm. Figure \ref{elbow} shows the line graph when 2-14 clusters are selected using the Elbow method, which can help find the elbow point. In K-means clustering, the potential elbow points are [3, 4, 6, 8]. In Birch clustering, [3, 9, 12] can be observed as elbow points. These cluster numbers will be used in our further experiments.

\subsection{Evaluation Metrics}

We employ accuracy, recall (also known as true positive rate), precision, FPR( also known as false positive rate), f1 score, and Receiver operating characteristic(ROC) as performance indicators. We use True Positive(TP) to identify correctly classified positive samples, False Negative(FN) for incorrectly classified positive samples, False Positive(FP) for incorrectly classified negative samples, and True Negative(TN) for correctly classified negative samples. Other measures using TP, FN, FP, and TN are as follows.
\begin{equation}\label{eq:ACC}
	Accuracy = \frac{TP+TN}{TP + TN + FP + FN}
\end{equation}

\begin{equation}\label{recall} Recall(True Positive Rate) = \frac {TP}{TP + FN}\end{equation}

\begin{equation} \label{precision}Precision = \frac {TP}{TP + FP}\end{equation}

\begin{equation} \label{FPR}FPR (False Positive Rate) = \frac {FP}{TN + FP}\end{equation}

\begin{equation} \label{f1}F1-score = 2\times \left ({\frac {Precision\times Recall}{Precision + Recall}}\right)\end{equation}

\begin{equation} \label{roc} AUC_{ROC}=\int _{0}^{1} \frac {TP}{TP+FN}d\frac {FP}{TN+FP}\end{equation}

\subsection{Results}

According to the experimental results in Table \ref{tab:k_result}, both k-means and Birch can achieve better performance than MLP alone. When using K-means with MLP, the best cluster number is 4, and it can get 99.70\% accuracy. When using Birch with MLP, the optimal cluster number is 12, which can achieve 99.73\% accuracy and is higher than using k-means.

\begin{figure}
\centering
\includegraphics[width=0.48\textwidth]{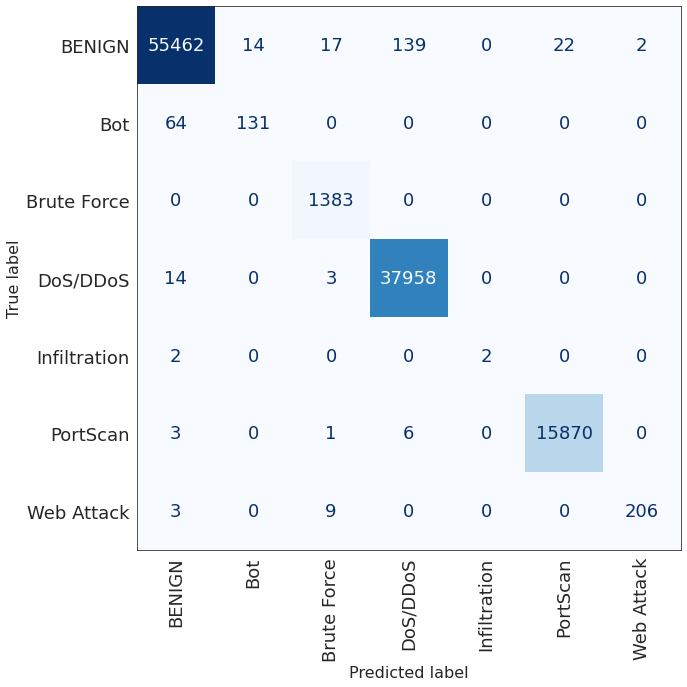}
\caption{Confusion Matrix}
\label{fig:confusion} 
\end{figure}

\begin{figure}
\centering
\includegraphics[width=0.48\textwidth]{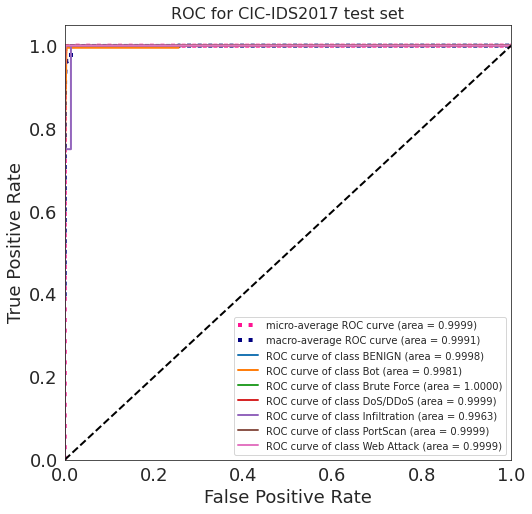}
\caption{ROC Curve}
\label{fig:ROC} 
\end{figure}

Figure \ref{fig:confusion} and figure \ref{fig:ROC} demonstrated the confusion matrix and ROC curve obtained by the model. As shown in Figure \ref{fig:confusion}, only a few samples are misclassified. In Figure \ref{fig:ROC}, the ROC of each class is plotted using the one versus all principle. The roc area of almost all classes reaches a score over 99\%, which means that the model has good generalization.

Table \ref{tab:result} presented the performance of each class using Birch with a cluster number of 12. Benign, Brute Force, DoS/DDoS and Web Attack achieved over 95\% accuracy. The performance of Bot and infiltration is slightly lower than other classes, which may be caused by their insufficient training samples. Furthermore, our model achieves a very low FPR of 0.15\%.

\subsection{Comparison}

In table \ref{tab:compare}, we compared the performance of our model with other studies using MLP as a classifier. It can be found that our model achieves the best performance in both f1 score and accuracy. Although the FPR is not the lowest, it is also a relatively good performance.

\section{Conclusion and Future Work}
\label{sec:con}
In this paper, we proposed a two-stage intrusion detection system that combines both an unsupervised clustering algorithm and a supervised machine learning model. In our method, we first use the Birch or k-means algorithm as an unsupervised clustering algorithm to group unlabeled data. The pseudo-label generated by the clustering algorithm is then added as an extra feature along with other features to our MLP model for multi-classification of network anomalies. In the experiment, we used the elbow method and information gain to determine the potential cluster numbers. The experimental results show that using Birch and K-Means in our approach can help improve the multi-classification performance of the MLP classifier. The best accuracy performance of 99.73\% can be obtained when using Birch with cluster number 12. After comparison, our model outperformed similar studies.

This paper initially explored the impact of two unsupervised clustering algorithms on MLP models. There are many variants of Birch and K-Means unsupervised clustering algorithms that can cluster unlabeled data more accurately. In the future, we plan to explore the effects of Birch and K-Means variants and other hyper-parameters of these algorithms on our hybrid model.

\bibliographystyle{IEEEtran}
\bibliography{mybib}

\end{document}